**Competing scanning tunneling microscope tip-interlayer interactions for twisted multilayer graphene on the a-plane SiC surface**

P. Xu [a], M.L. Ackerman [a], S.D. Barber [a], J.K. Schoelz [a], P.M. Thibado [a,*], V.D. Wheeler [b], L.O. Nyakiti [b], R.L. Myers-Ward [b], C.R. Eddy, Jr. [b], D.K. Gaskill [b]

[a] *Department of Physics, University of Arkansas, Fayetteville, AR 72701, USA*

[b] *U.S. Naval Research Laboratory, Washington, DC 20375, USA*

**ABSTRACT**

Scanning tunneling microscopy (STM) images are obtained for the first time on few layer and twisted multilayer epitaxial graphene states synthesized on n$^+$ 6H-SiC a-plane non-polar surface. The twisted graphene is determined to have a rotation angle of 5.4° between the top two layers, by comparing moiré patterns from stick and ball models of bilayer graphene to experimentally obtained images. Furthermore, the experimental moiré pattern shows dynamic behavior, continuously shuffling between two stable surface arrangements one bond length apart. The moiré pattern shifts by more than 1 nm, making it easy to observe with STM. Explanation of this dynamic behavior is attributed to electrostatic interactions between the STM tip and the graphene sample.

* Corresponding author.
*E-mail address:* thibado@uark.edu (P.M. Thibado).

# 1. Introduction

The novel electronic properties of graphene have been spurring intense research interest ever since this two-dimensional (2D) material was first successfully isolated [1]. These intriguing properties, such as ballistic transport [2], the quantum Hall effect [3], and ultra-high mobility [4], mark graphene as a potentially crucial material in post-silicon electronics. In particular, epitaxial graphene grown on SiC via thermal decomposition has been identified as one of the most likely avenues to graphene-based electronics [5]. As a highly resistive material, semi-insulating SiC has a minimal effect on current flow in graphene, and it is already available in the form of large-diameter wafers compatible with current industrial technology [6-11]. This enormous promise has, as a result, triggered a much closer investigation of microscopic structural properties of epitaxial graphene on SiC [12-14]. For example, large-scale atomic force microscopy images have revealed 1-2 nm high ridges running parallel and perpendicular to steps in the substrate [15; 16]. These have typically been attributed to stress induced in the graphene layers during post-growth cooling as graphene expands and the substrate contracts [17-20]. Atomic-scale scanning tunneling microscopy (STM) experiments performed on the ridges established them as buckled regions of graphene and showed that they occasionally rearrange under the influence of the STM tip [21; 22].

However, the lack of an intrinsic band gap in graphene has proved a major deterrent to its application in digital electronic devices. One possible solution to this problem is graphene nanoribbons, which have a band gap inversely proportional to their width due to quantum confinement [23]. Initial methods for their manufacture involved cutting ribbons from larger sheets, resulting in disordered edges detrimental to transport [24]. Recently, though, graphene nanoribbons as narrow as 40 nm have been grown directly on nanofacets in SiC surfaces, advantageously preserving their structural integrity [25]. This discovery has spurred the interest in studying other surface orientations of SiC.

Another approach to introducing a band gap is by using bilayer graphene, which has a band gap in proportion to the strength of an applied perpendicular electric field [26; 27]. Bilayer graphene may also be grown directly on SiC surfaces [28], but its electronic properties will depend strongly on the angle of rotation between the two layers [29]. For instance, a reduction in the carrier velocity as a function of twist angle has been predicted [30], although this effect has recently been contested for small (< 3°) twist angles [31]. It has also been predicted that twisted bilayer graphene will have no band gap even in the presence of a perpendicular field [32]. It is therefore imperative to have more experimental studies of the properties of bilayer graphene [33].

In this paper, we study epitaxial graphene grown on the non-traditional, non-polar 6H-SiC($11\bar{2}0$) or a-plane surface and show that there are regions of few layer and twisted multilayer graphene. Experimental STM images are presented which display the characteristic moiré pattern corresponding to a 5.4° twist angle of the top layer relative to the layer underneath. In addition, we show that interactions between the tip and sample cause the top layer to undergo small horizontal displacements, with a ~10x larger displacement in the moiré pattern.

## 2. Experimental

The epitaxial graphene sample used in this study was grown on the non-polar face of a n$^+$ 6H-SiC a-plane nominally-oriented sample (Aymont Technology) measuring 16 mm × 16 mm. Growth was carried out in a commercial hot-wall Aixtron VP508 chemical vapor deposition reactor. Prior to graphene growth, the SiC substrate was etched *in situ* in a 100 mbar H$_2$ ambient at 1520 °C for 50 min. This etching produces a controlled starting surface that is smooth, specular and has an atomic force microscopy RMS roughness values of 0.12 nm (10 μm x10 μm height scan) that are within instrument resolution limit (Veeco D3100 operated in tapping mode), as shown in Fig. 1(a). After the $H_2$ etching step, the ambient was switched to Ar with a transition time of 2 minutes during which pressures varied by ±50% around 100 mbar, this was followed by a temperature ramp to 1620 °C. The subsequent 60 minute graphene synthesis process was conducted under a flowing Ar environment of 20 standard L/min at 100 mbar, with a substrate temperature of 1620 °C. After growth, the sample was cooled in Ar to room temperature, diced into 5 mm × 5 mm samples, and diamond scribed labels were added to the back side. The post-growth morphology of the sample has terraces and steps throughout the surface as shown in Fig. 1(b). Next, Raman data was collected for the sample using a Thermo DXR system. A 532 nm, 8 mW laser was used as the pump probe with a spot size 0.7 μm. After characterization with Raman, the sample was secured, sealed, and sent to the STM facility.

Constant-current filled-state STM images were obtained using an Omicron ultrahigh-vacuum (base pressure is $10^{-10}$ mbar), low-temperature model STM operated at room temperature. The sample was mounted with silver paint onto a flat tantalum sample plate and transferred through a load-lock into the STM chamber where it was electrically grounded. STM tips were electrochemically etched from 0.25 mm diameter polycrystalline tungsten wire via a custom double lamella method with an automatic gravity-switch cutoff. After etching, the tips were gently rinsed with distilled water, briefly dipped in a concentrated hydrofluoric acid solution to remove surface oxides, and then transferred into the STM chamber. All images were acquired using a positive tip bias of 0.100 V and a tunneling current setpoint of 1.00 nA.

## 3. Results and discussion

The average roughness of the surface after graphene formation was determined from AFM to be 1.06 nm and different graphene regions bounded by wrinkles/ridges were found as shown in Fig. 1(c). The average size of the regions is around 1 micron. Confocal Raman spectrometry confirms the presence and layer thickness of epitaxial graphene on the surface as shown in Fig. 1(d). The intensity of major Raman modes (D, G and 2D) as a function of Raman shift after subtracting the SiC background are shown. The small D peak intensity indicates the absence of a significant number of defects. The 2D peak is symmetric with a full width at half maximum (FWHM) of ~50 cm$^{-1}$, which indicates that there are 2-4 layers of graphene on the surface of our sample. Note, a symmetric 2D peak having a FWHM of ~25 cm$^{-1}$ would indicate a monolayer [34-37].

Three STM images of the graphene/SiC a-plane surface are shown in Fig. 2. A large-scale (200 nm × 200 nm) image is first given in Fig. 2(a), and it is characterized by a series of terraces with the surface height increasing from left to right. A horizontal line profile spanning the STM image and extracted from near the bottom is displayed beneath it, showing the step height and width of the substrate terraces. The steps average approximately 2.5 nm in height, each followed by a plateau around 50 nm in width, which gives an overall nominal wafer miscut of 2-3º, which is also confirmed with larger scale AFM images. Atomically resolved STM images were obtainable virtually everywhere on the terraces, and two such small-scale images, measuring 6 nm × 4 nm, are presented in Figs. 1(b) and 1(c). The triangular atomic symmetry indicative of highly coupled, few layer Bernal stacked graphene[38], appears throughout Fig. 2(b), yet a honeycomb structure, characteristic of graphene that has been largely electronically decoupled [39] appears throughout Fig. 2(c). Note, however, that the topography of the latter also exhibits long-wavelength undulations. These two atomic-scale STM images show that few graphene layers were formed, which is not uncommon for thermal decomposition of SiC [40-42].

Further analysis of the modulated hexagonal pattern detected in Fig. 2(c) is performed in Fig. 3. First, a similar 6 nm × 6 nm atomic-scale STM image of the graphene/SiC surface displaying the honeycomb lattice with large alternating bright and dark spots is shown in Fig. 3(a). Then to highlight the topographic structure of the surface, two line profiles were extracted from the image and plotted in Fig. 3(b). The starting heights have been offset in order to see the detail in both curves. The upper (blue) curve was taken from top to bottom along the nearly vertical arrow in Fig. 3(a), and the lower (green) curve was taken from left to right along the nearly horizontal arrow. These directions were chosen to align with and bisect the bright spots, resulting in curves with peak-to-peak amplitude of 0.08 nm. The upper line has two peaks roughly 3 nm apart, whereas the two peaks of the lower line are separated by 4.5 nm. Using larger scale STM images with ten repeats the distance between the features is on average 2.8 nm and 4.7 nm, respectively.

In an attempt to replicate the peak separation distances found in the experimental image, two computer-generated 2D hexagonal lattices were stacked on top of one another, as depicted in Fig. 3(c). The top layer was rotated clockwise relative to the lower layer in increments of 0.1° around a normal axis running through a carbon atom approximately in the middle of Fig. 3(c) until the spatial periodicity matched that of Fig. 3(a). An angular displacement of 5.4° was found to best match the experimental results. This displacement is illustrated more clearly in Fig. 3(d), which shows a close-up of hexagons from the top and bottom layers around the point of rotation as cropped from Fig. 3(c).

The simulated model suggests that the STM image of Fig. 3(a) portrays a moiré pattern of twisted graphene with a twist angle of 5.4°. The twist angle between the top two layers causes carbon atoms in the top layer to precisely overlap with carbon atoms in the lower layer (i.e., AA stacked) in certain sections. It is known that AA stacking of graphene layers is less energetically stable than AB stacking [43]. This produces a local displacement between the two layers and subsequent height increase of the top layer, which is measured in the line profiles of Fig. 3(b). In fact, at this twist angle Luican *et al.* indicate significant decoupling compared to untwisted bilayer graphene [30], consistent with our experimental results showing that the honeycomb structure of monolayer graphene is visible everywhere in Fig. 3(a) [39].

Interestingly, the experimental moiré pattern in many cases exhibited unusual systematic time dependence. To illustrate this effect, a series of 6 nm × 6 nm STM images was continuously collected without delay between scans at a single position on the graphene/SiC surface, where each scan took ~2 min. These images are presented in sequential order in Fig. 4, and overlaid on each is a diamond-shaped box outlining the unit cell of the moiré periodicity. Between Fig. 4(a) and Fig. 4(b), the cell has shifted up and to the right by about 1 nm. In the image taken immediately afterward [Fig. 4(c)], the cell has returned to its original position, yet in Fig. 4(d) it has shifted again as in Fig. 4(b). Finally, the initial configuration is once more observed in Fig. 4(e). Comparing the positions of the moiré pattern's unit cell in this series of images implies that there are two distinct configurations occurring between the top and bottom graphene layers, with movement between them. Previous STM experiments on graphene and graphite have established that an electrostatic interaction between the tip and the sample can cause small horizontal shifts in the top graphitic layer [43], and here the moiré pattern images serve to magnify the fidelity of the movement. Since its spatial periodicity is approximately 10 times that of monolayer graphene, the displacement in the moiré pattern is an order of magnitude larger than the actual top layer displacement.

In order to fully characterize the bi-stable graphene configurations, as seen in Fig. 4, the 5.4° twisted graphene model was again employed. First, the simulated graphene bilayer was cropped to match the dimensions and moiré pattern of the STM image in Fig. 4(a). The resulting image with the moiré unit cell outlined is shown in Fig. 5(a). Then the top layer in the model was shifted by various amounts and in various directions. The resulting images were next compared to Fig. 4(b), and it was found that a shift along the carbon-carbon bond axis by one bond length best replicated the experimental results. This configuration is shown in Fig. 5(b). In order to more clearly see the atomic-scale shift, the two model images were cropped down to a few graphene unit cells. The small-scale picture before the shift is shown in Fig. 5(c), below the corresponding large-scale picture. Near the lower left corner the two layers exhibit AB stacking. The small-scale image after the horizontal shift is shown in Fig. 5(d), below its corresponding large-scale image. Near the lower left corner the two layers now exhibit AA stacking, indicative of a shift by one full bond length. The chosen magnitude and direction of the shift given above corresponds to transitions between the stacking patterns of graphite layers, as well [44]. Interestingly, shifting the top layer of trilayer graphite by this amount and in this direction shifts the stack from ABA stacking to ABC stacking.

After this analysis of the local structural changes that occur between the two different moiré pattern configurations, we developed a proposed mechanism for the process. As the STM tip scans over the sample, the tip is disproportionately electrostatically attracted to the nearby AA stacked regions, which have a larger charge density and height. The STM tip, when in the correct position naturally exerts a small horizontal force, pulling the AA stacked graphene into the energetically favored AB stacking position and, thereby leading to a decrease in the energy of that part of the system. However, the areas of AB stacked graphene are consequently pushed into AA stacking, increasing the energy in those parts of the system. So the total energy of the top two layers remains the same, even as the moiré pattern has shifted by a nanometer, as in Fig. 4. Thus, the combination of unfavorable energetics near AA stacked twisted graphene, along with electrostatic interactions between the tip and the sample, results in the formation of a dynamic moiré pattern with two stable states.

## 4. Conclusions

Atomic-scale STM images of epitaxial graphene grown via thermal decomposition on the non-traditional, non-polar 6H-SiC a-plane surface were acquired for the first time. Multiple arrangements of few layer graphene were observed on the surface, including twisted multilayer graphene. Using the spatial periodicity of the associated moiré pattern, an angle of 5.4° between the top two layers was determined for the twisted graphene. Dynamic atomic-scale horizontal displacements in the twisted graphene were concluded based on magnified bimodal behavior in the moiré pattern, likely facilitated by partial decoupling between the graphene layers and the electrostatic influence of the STM tip.

**Acknowledgments**

P.X. and P.M.T. gratefully acknowledge the financial support of ONR under grant N00014-10-1-0181 and NSF under grant DMR-0855358. Work at the U.S. Naval Research Laboratory is supported by the Office of Naval Research. L.O.N. gratefully acknowledges postdoctoral fellowship support through the ASEE.

**Figure captions**

**Fig. 1** (a) An AFM image of the SiC(11$\bar{2}$0) or a-plane surface after it was etched *in situ* in a 100 mbar H$_2$ ambient at 1520 °C for 50 min. (b) An AFM image of the sample surface after multilayer epitaxial graphene was formed. (c) A zoomed-in view AFM image of multilayer epitaxial graphene. (d) Raman spectra of the surface obtained using confocal Raman spectrometry with an excitation wavelength of 532 nm and a laser spot size of 0.7 µm.

**Fig. 2** (a) Large-scale (200 nm × 200 nm) STM image of epitaxial graphene on 6H-SiC a-plane surface showing a terraced structure. Line profile beneath was extracted horizontally across the image from near the bottom. (b) Atomic-scale (6 nm × 4 nm) STM image taken from a terrace and having the appearance of few layer graphene. (c) Atomic-scale (6 nm × 4 nm) STM image taken from a different region and having the appearance of graphene.

**Fig. 3** (a) Experimental 6 nm × 6 nm STM image of epitaxial graphene on 6H-SiC a-plane surface displaying a moiré pattern. (b) Line profiles taken along the arrows shown in (a). (c) Twisted bilayer graphene model. The top layer has been rotated 5.4° relative to the bottom layer. (d) Small-scale image of the model shown in (c). The axis of rotation is in the lower left corner.

**Fig. 4** Chronological sequence of atomic-scale 6 nm × 6 nm STM images of epitaxial graphene on 6H-SiC a-plane surface taken at a single location and revealing the horizontal movement of the top layer of graphene. A diamond representing the unit cell of the moiré pattern has been drawn on each image, and two configurations are found to exist.

**Fig. 5** (a) Twisted bilayer graphene model. The top layer is rotated 5.4° relative to the bottom layer. The image is cropped so that the regions of high carbon-carbon overlap match the positions of the bright spots in Fig. 4(a). (b) The same model after the top layer has been shifted by one bond length. The regions of high carbon-carbon overlap now match the positions of the bright spots in Fig. 4(b). (c,d) Small-scale, zoomed-in images of the models shown in (a,b), respectively.

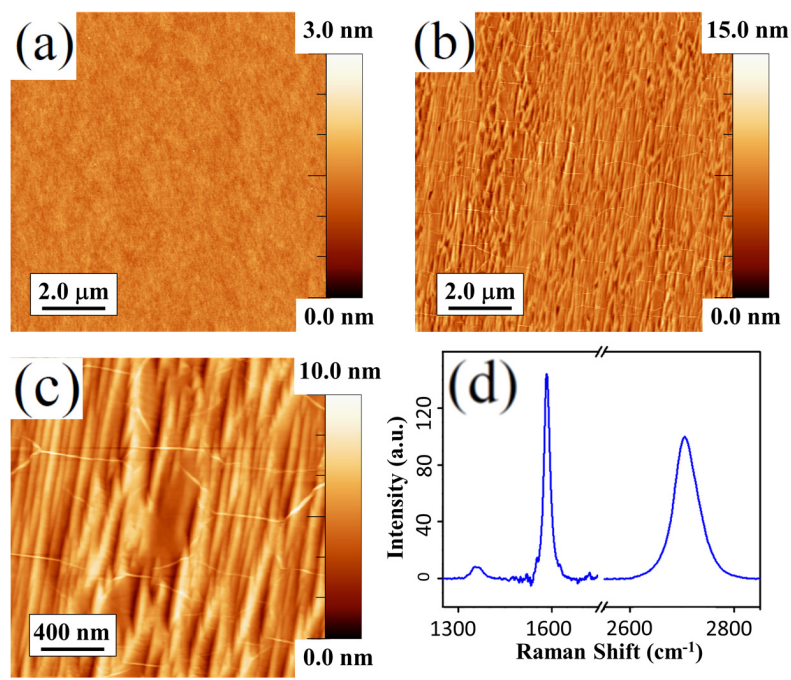

Fig. 1

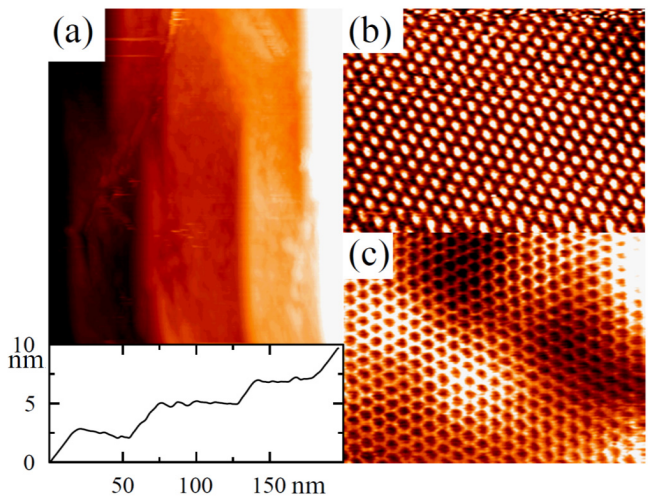

Fig. 2

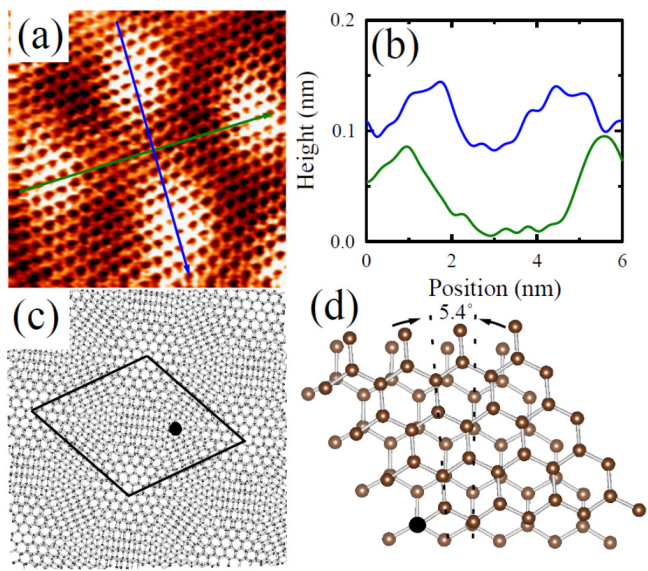

Fig. 3

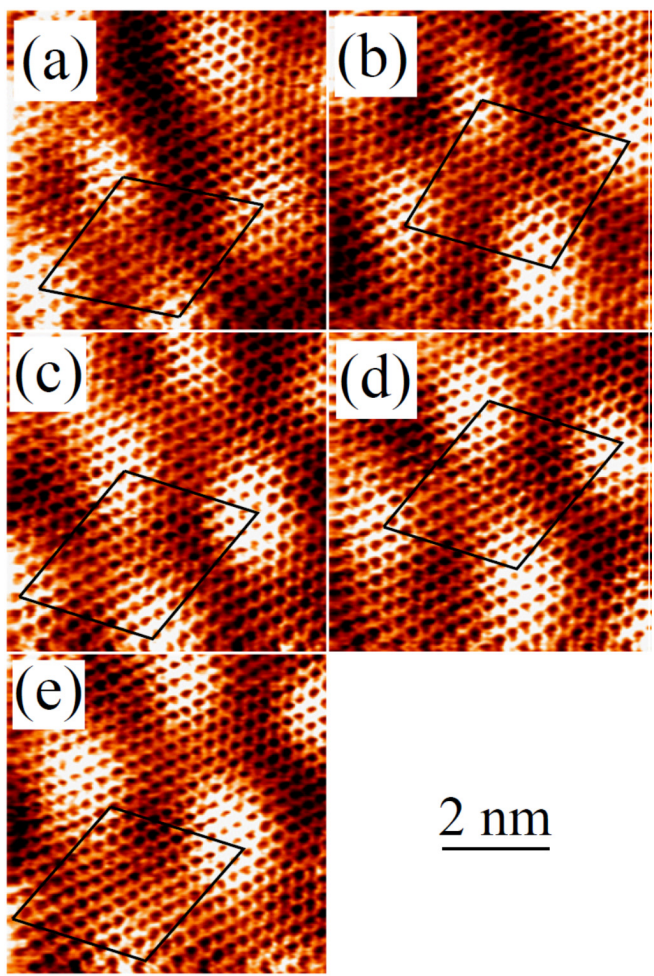

Fig. 4

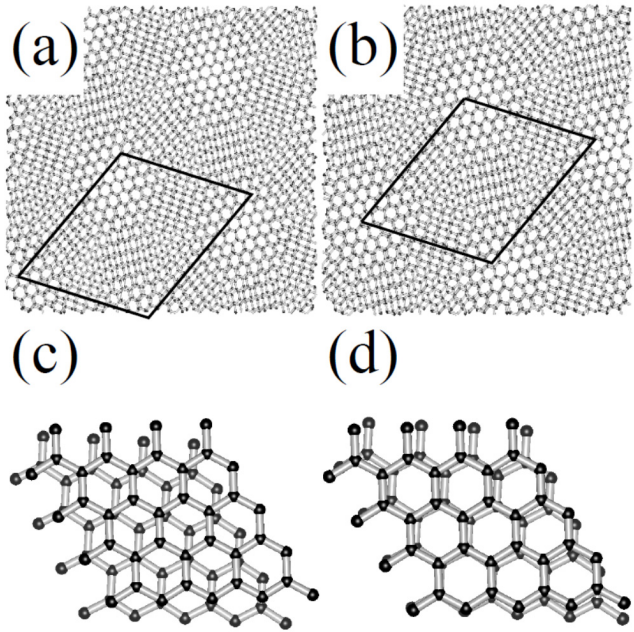

Fig. 5